\documentclass[letterpaper]{article} 
\usepackage{aaai2026}  
\usepackage{times}  
\usepackage{helvet}  
\usepackage{courier}  
\usepackage[hyphens]{url}  
\usepackage{graphicx} 
\usepackage{natbib}  
\usepackage{caption} 
\usepackage{algorithm}
\usepackage{algorithmic}
\usepackage{amsmath}
\usepackage{newfloat}
\usepackage{listings}
\usepackage{booktabs}
\DeclareCaptionStyle{ruled}{labelfont=normalfont,labelsep=colon,strut=off} 
\floatstyle{ruled}
\newfloat{listing}{tb}{lst}{}
\floatname{listing}{Listing}
\title{DMGIN: How Multimodal LLMs Enhance Large Recommendation Models for Lifelong User Post-click Behaviors}
\author{
    Zhuoxing Wei\textsuperscript{\rm 12}\equalcontrib,
    Qingchen Xie\textsuperscript{\rm 13}\equalcontrib,Qi Liu\textsuperscript{\rm 4}\thanks{Corresponding author}\\
}
\affiliations{
    \textsuperscript{\rm 1}The core local commerce segment of Meituan\\
    \textsuperscript{\rm 2}PeKing University\\
    \textsuperscript{\rm 3}Tsinghua University\\
    \textsuperscript{\rm 4}University of Science and Technology\\
    weizhuoxing@meituan.com, xieqingchen@meituan.com, qiliu67@mail.ustc.edu.cn

}

\usepackage{bibentry}

\begin{document}

\maketitle

\begin{abstract}
Modeling user interest based on lifelong user behavior sequences is crucial for enhancing Click-Through Rate (CTR) prediction. However, long post-click behavior sequences themselves pose severe performance issues: the sheer volume of data leads to high computational costs and inefficiencies in model training and inference. Traditional methods address this by introducing two-stage approaches, but this compromises model effectiveness due to incomplete utilization of the full sequence context. More importantly, integrating multimodal embeddings into existing large recommendation models (LRM) presents significant challenges: These embeddings often exacerbate computational burdens and mismatch with LRM architectures. 
To address these issues and enhance the model's efficiency and accuracy, we introduce Deep Multimodal Group Interest Network (DMGIN). Given the observation that user post-click behavior sequences contain a large number of repeated items with varying behaviors and timestamps, DMGIN employs Multimodal LLMs(MLLM) for grouping to reorganize complete lifelong post-click behavior sequences more effectively, with almost no additional computational overhead, as opposed to directly introducing multimodal embeddings. To mitigate the potential information loss from grouping, we have implemented two key strategies. First, we analyze behaviors within each group using both interest statistics and intra-group transformers to capture group traits. Second, apply inter-group transformers to temporally ordered groups to capture the evolution of user group interests. Our extensive experiments on both industrial and public datasets confirm the effectiveness and efficiency of DMGIN. The A/B test in our LBS advertising system shows that DMGIN improves CTR by 4.7\% and Revenue per Mile by 2.3\%.

\end{abstract}


\section{Introduction}
\begin{figure}
    \centering
    \includegraphics[width=\linewidth]{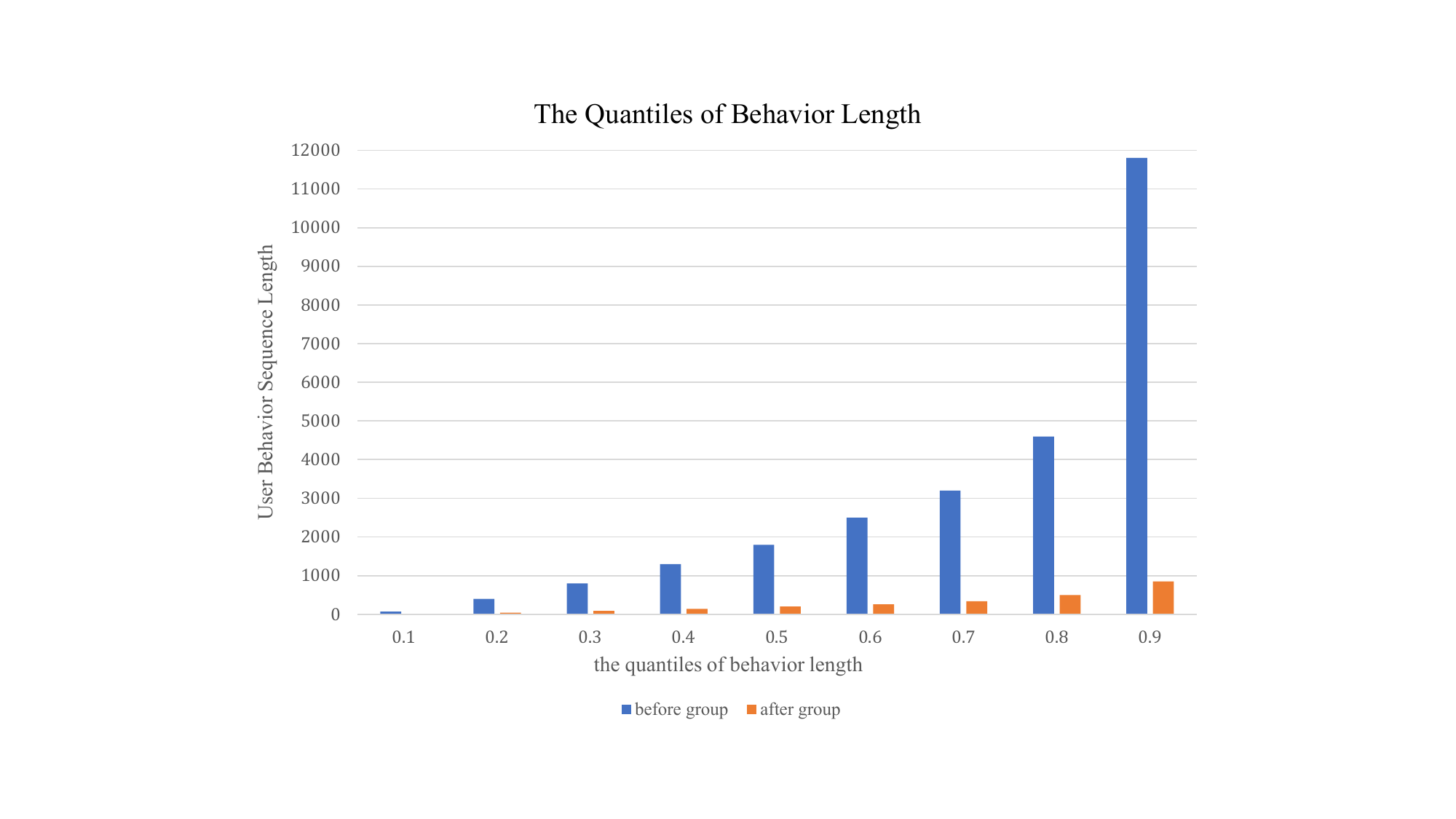}
     \vspace{-1em}
    \caption{The distribution of behaviors' length affected by grouping on \emph{item\_id}. We rank samples on behaviors' length in ascending order. The left half shows the corresponding behavior's length at different quantiles. The right half displays the distribution of the behavior quantity of each group.}
    \label{fig:motivation_num_hist}
    \vspace{-1em}
\end{figure}
The prediction of Click-Through Rate (CTR) \cite{guo2017deepfm, zhou2018deep, zhou2019deep} plays a pivotal role in online advertising and recommendation systems, directly influencing user experience and revenue generation. Take the booming food delivery industry as an example: merchants strive to make their offerings appealing by crafting enticing descriptions and accompanying them with mouthwatering photos and videos. This prompts users to repeatedly compare various food options, scrutinizing text descriptions, weighing customer reviews, and evaluating visual content such as dish photos, generating a wealth of behavioral data. This includes a spectrum of post-click interactions, such as diving into detailed menus, analyzing user ratings, or adding items to virtual carts. For digital platforms, amid the exponential growth of such user post-click behaviors and multimodal contextual information (e.g., text descriptions, images, and videos), accurately deciphering user intent and interests has become crucial for delivering personalized content. A key challenge in CTR prediction is the effective modeling of user post-click behaviors alongside multimodal content, especially when such behaviors unfold over extended timeframes, generating elaborate and intricate interaction sequences. Existing methods have generally treated long behavioral sequences \cite{pi2020search, chang2023twin} and multimodal content \cite{sheng2024enhancing, luo2024qarm} as separate entities for modeling, thereby failing to capture their intrinsic interdependencies.

From the perspective of long sequences, modeling user interests based on lifelong behavioral sequences is crucial for improving CTR prediction. Traditional methods tackle this by adopting two-stage approaches (e.g., first retrieving relevant subsequences and then applying attention mechanisms). SIM~\cite{pi2020search} applies the item's category as a relatedness metric to retrieve items. TWIN~\cite{chang2023twin} solves the distribution discrepancy by make two stages share an efficient target attention network. Following research change to retrieve relevant clusters rather than original items, preserving more information about users' long-term preferences. TWIN V2~\cite{si2024twin} employs a hierarchical clustering method to group items with similar characteristics into a cluster. HSTU\cite{zhai2024actions} focuses on incorporating actions into long sequences and reformulates recommendation problems as sequential transduction tasks within a generative modeling framework. incorporates global tokens into the Transformer \cite{vaswani2017attention} module to enhance performance and adopts a token-merging strategy to reduce the computational overhead of the Transformer. Yet, long post-click behavior sequences themselves introduce severe performance bottlenecks and increase problem complexity. A key observation here is that user post-click behavior sequences contain a large number of repeated items, accompanied by varying associated behaviors and timestamps. Existing methods often compromise model effectiveness due to their incomplete utilization of full sequence contexts; meanwhile, efforts to reduce sequence length by removing duplicates further exacerbate information loss.

Alongside the remarkable evolution of Multimodal Large Language Models (MLLM), researchers in recommendation systems have increasingly recognized the potential of multimodal information for modeling user interests. For instance, the Prefix-ngram method \cite{zheng2025enhancing} introduced a novel token parameterization technique termed Semantic ID prefix ngram, which scales up codebook values—an innovation that significantly improves the performance of the original Semantic ID approach. Similarly, QARM \cite{luo2024qarm} proposed an item alignment mechanism that generates consistent multimodal representations, transforming raw multimodal features to align with the actual distribution of user-item interactions for downstream business tasks. BBQRec \cite{li2025bbqrec} incorporated a behavior-semantic alignment module to disentangle modality-agnostic behavioral patterns from noisy modality-specific features, ensuring that semantic IDs are inherently tied to recommendation objectives. Additionally, Letter \cite{wang2024learnable} integrated a contrastive alignment loss for collaborative regularization and a diversity loss to mitigate code assignment bias, thereby enhancing the diversity of token assignments, alleviating item generation bias, and ensuring fairness in item generation.
However, existing applications of MLLM in recommendation systems \cite{liu2025llm} have notable limitations. Directly integrating multimodal embeddings into LRM \cite{chai2025longer} poses significant challenges: these embeddings often increase computational burdens and suffer from architectural mismatches with LRM, hindering their seamless integration and effective utilization in recommendation scenarios. Notably, they fail to capture lengthy lifelong post-click behaviors. Thus, how to efficiently model the application of MLLM in LRM to leverage lifelong user post-click behaviors has become an urgent issue.

To address these issues and enhance the model’s efficiency and accuracy, we propose the Deep Multimodal Group Interest Network (DMGIN). While user behavior sequences contain a large volume of post-click interactions, we observe that these sequences include numerous repeated or highly relevant shops—identifiable via consistent shop names and images—accompanied by varying behavioral details and timestamps. We hypothesize that users’ interest in a shop is shaped by the descriptions and images they encounter. Leveraging this insight, DMGIN utilizes MLLM to more effectively reorganize complete lifelong post-click behavior sequences by grouping these highly relevant shops into unified clusters. First, to train the MLLM component, we pre-train a CLIP-like model using diverse multimodal pairs (e.g., shop name and shop images, food name and food description), helping it learn to link text and visuals of the same entity. Second, during inference, we generate multimodal embeddings for all shops (not each user behavior) to avoid heavy computation from billions of daily interactions. Instead of directly using these embeddings, DMGIN clusters shops with K-means based on their embeddings. We check cluster balance (ensuring even distribution) and visualize shop embeddings for human review to ensure reasonable grouping. In short, this adds almost no extra computation. It reorganizes users’ lifelong post-click behaviors into interest groups, slashing sequence length from tens of thousands to hundreds while keeping key preferences intact.

Notably, DMGIN groups users' lifelong post-click behaviors, successfully reducing the length of user sequences by tens to hundreds of times. However, this process inevitably leads to performance degradation, as the original sequence structure is disrupted. To mitigate the potential information loss caused by grouping, we have implemented two key novel strategies. First, we analyze behaviors within each group using interest statistics to capture statistical information, which includes counters for each behavior type as well as maximum and average time pooling within each group. Second, DMGIN employs intragroup transformers to model varying actions and timestamps, capturing the interest evolution of each group, specifically, enabling fine-grained capture of user interests within a group.

After extracting the group representations, we apply multi-layer inter-group transformers to temporally ordered groups to capture the evolutionary patterns of users' group-based interests. This component, which is also a core part of industrial long-range models (LRM), is crucial as it enables the generation of fully cross-referenced group representations. Finally, we refine the target attention mechanism by calculating the attention between target items and groups to identify candidate-specific interests.

Overall, we make the following contributions:
\begin{itemize}
\item We propose the Deep Multimodal Group Interest Network (DMGIN), a unified framework for modeling long behavioral sequences with multimodal embeddings in industrial LRM.
\item We leverage MLLM to generate multimodal representations for shops and apply K-means clustering to all shop entities.
\item We introduce a grouping method to process lengthy user post-click sequences, accompanied by two key strategies to enhance group representations.
\item Real-World Application and A/B Testing: Our DMGIN approach has been successfully deployed in a large-scale Location-Based Services (LBS) advertising system. A/B test results demonstrate that DMGIN improves click-through rate (CTR) by 4.5\% and Revenue per Mile (RPM) by 2.0\%, validating its practical effectiveness in real-world scenarios.
\end{itemize}

\begin{figure*}
    \centering    
    \includegraphics[width=0.9\linewidth]{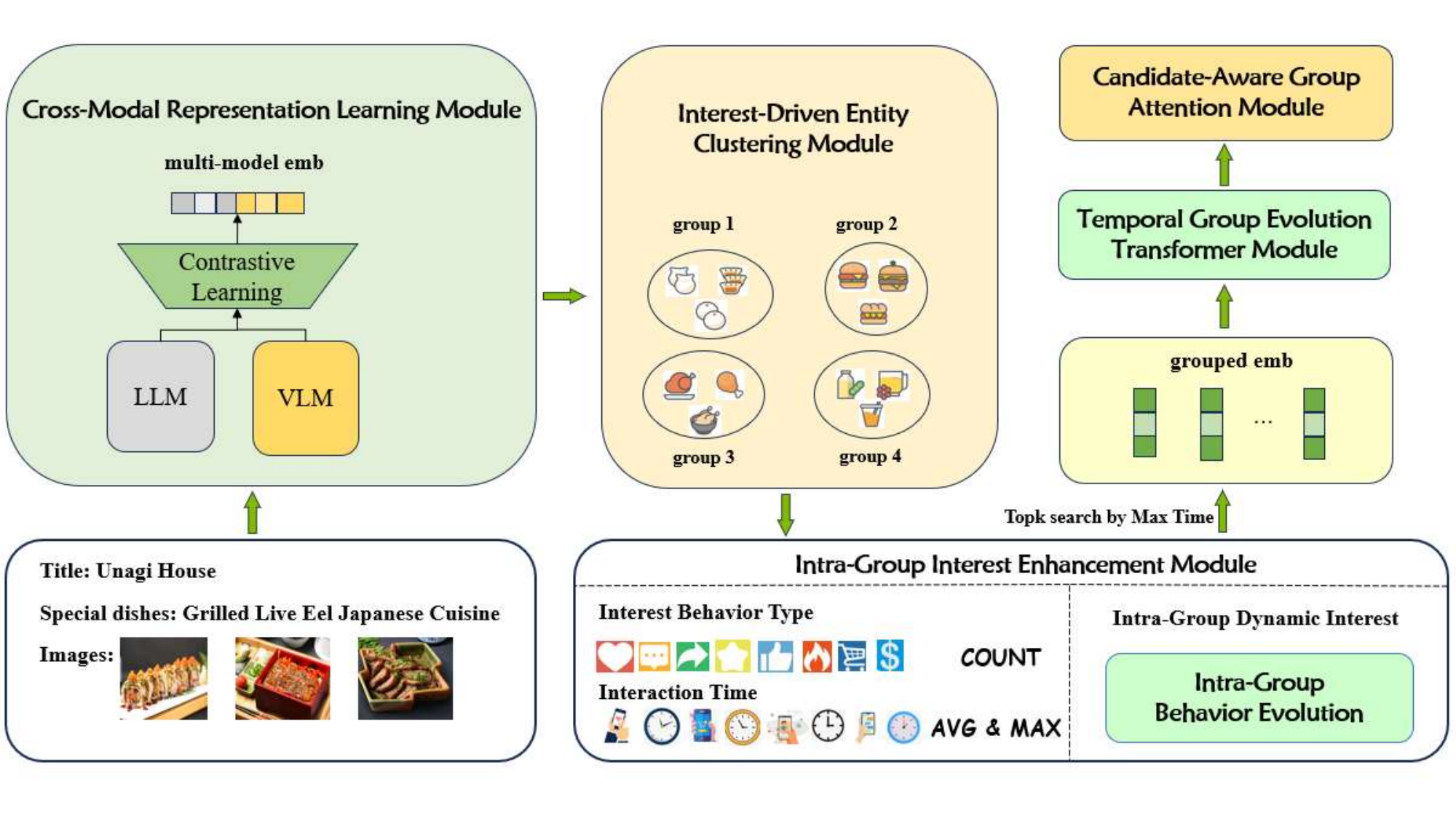}
    \caption{The overall framework of Deep Multimodal Group Interest Network (DMGIN). DMGIN consists of Cross-Modal Representation Learning Module (CMRLM), Interest-Driven Entity Clustering Module(IDECM), Intra-Group Interest Enhancement Module(IGIEM), Temporal Group Evolution Transformer(TGET) and Candidate-Aware Group Attention Module (CAGAM).}
    \label{fig:dmgin}
\end{figure*}
\section{Method}
We propose the Deep Multimodal Group Interest Network (DMGIN). A key observation guides our approach: although user behavior sequences contain massive post-click interactions, they are characterized by numerous repeated or semantically related shops—entities distinguishable through consistent identifiers (e.g., shop names) and visual cues (e.g., images)—accompanied by varying behavioral details (e.g., interaction timestamps). We hypothesize that users’ interest in a shop is fundamentally shaped by the multimodal information they encounter, such as textual descriptions and visual content. Leveraging this insight, DMGIN employs a Multimodal Large Language Model (MLLM) to reorganize users’ complete lifelong post-click sequences by grouping highly relevant shops into unified clusters, thereby achieving efficient and effective sequence modeling.

\subsection{Cross-Modal Representation Learning Module}
The Cross-Modal Representation Learning Module (CMRLM) is designed to learn robust, semantically aligned multimodal representations of shops, laying the foundation for meaningful clustering.
To train this module, we develop a CLIP-like dual-tower model pre-trained on a diverse dataset of entity-centric multimodal pairs. These pairs include:
(shop name, shop images) to align textual identifiers with visual appearances;
(food\/category name, product descriptions) to link category labels with semantic attributes;
(business keywords, scene images) (e.g., "pet-friendly café" paired with interior images) to connect contextual text with visual scenarios.
The pre-training objective focuses on maximizing the similarity between embeddings of matching text and visual inputs while minimizing similarity for mismatched pairs. This ensures that the learned representations reside in a shared latent space where semantically equivalent entities (regardless of modality) are positioned closely, enabling reliable cross-modal comparisons—critical for subsequent clustering.
\subsection{Interest-Driven Entity Clustering Module}
Building on the cross-modal embeddings generated by the CMRLM, the Interest-Driven Entity Clustering Module(IDECM) groups shops into semantically coherent clusters to achieve efficient sequence compression.
During the inference phase, this module first generates multimodal embeddings for all unique shops in the system (rather than processing individual user behaviors). This design avoids the computational explosion that would result from encoding billions of daily user interactions, as shop-level embeddings are computed once and reused across all users.
Instead of directly feeding these embeddings into the recommendation pipeline, the module applies K-means clustering to group shops based on their embedding similarity. To ensure clustering quality, two validation steps are implemented:
Cluster Load Balance Check: We analyze the distribution of shops across clusters to prevent extreme imbalance (e.g., overly large or small clusters) that could dilute interest signals.
Human Evaluation via Visualization: We visualize the latent space distribution of shop embeddings (using techniques like t-SNE) to manually verify that semantically similar shops (e.g., same cuisine types, brand chains) are grouped together.

The output of this module is a mapping from individual shops to interest clusters, which facilitates the transformation of raw user behavior sequences: tens of thousands of discrete post-click interactions are reorganized into hundreds of cluster-level interest units. This compression achieves a drastic reduction in sequence length while retaining core long-term user preferences, with computational overhead that remains negligible compared to utilizing raw multimodal embeddings.

\subsection{Intra-Group Interest Enhancement Module}
Figure~\ref{fig:dmgin} details the Intra-Group Interest Enhancement Module (IGIEM), with its right-bottom section illustrating the module's core functionality: extracting enhanced interest representations for user groups.
Notably, while DMGIN effectively groups users' lifelong post-click behaviors, achieving a sequence-length reduction of tens to hundreds of times, this compression inevitably disrupts the original sequence structure, potentially degrading model performance. To address such information loss introduced by grouping, we propose two novel strategies.
\subsubsection{Group-Level Interest Statistics}
Given the large number of post-click behavior types, we categorize these fine-grained types into several interest-based categories, including strong interest, weak interest, negative feedback, and payment-related actions/amounts, among others. Within each group, we count the total occurrences of each categorized behavior type. From a temporal perspective, we compute the average and maximum duration within a group. Additionally, we derive the average consumption amount for all purchase-related behaviors. These three dimensions, quantity, time, and monetary value, serve as direct indicators of user interest intensity. All such statistical features are preprocessed during the offline data processing phase, incurring no additional inference-time overhead. We represent the Interest Statistics as Eq.~(\ref{eq:stat_attr}).
\begin{equation}
    \label{eq:stat_attr}
    \mathbf{stat}_s=[\mathbf{stat}_{counts},\mathbf{stat}_{max\_time},...,\mathbf{stat}_{avg\_price}]
\end{equation}
where $\mathbf{stat_s}$ is the interest statistics of the $s$-th group.

\subsubsection{Intra-Group Behavior Evolution}
While Group-Level Interest Statistics effectively capture interest intensity, they fail to characterize intra-group interest evolution. Within each group, spatiotemporal attributes, such as timestamps and locations of behaviors, can reflect the evolutionary trajectory of coarse-grained interests of users. Meanwhile, heterogeneous interactions reveal users’ dynamic attitudes toward items: for instance, recent clicks may strongly influence current bookmarking or cart-adding actions, while older clicks have minimal impact, yet early purchases might exert greater influence on current clicks.
To address this, we introduce sequential attributes—including timestamps, locations, and behavior types—to supplement information about interest dynamics. We then employ intra-group transformers to model these varying timestamps, locations and actions, enabling the capture of each group’s interest evolution and, more specifically, fine-grained user interests within the group.
Suppose there are maximum of $B$ behaviors in each group as $\mathbf{b}=<\mathbf{b}_{1}, \mathbf{b}_{2}, ..., \mathbf{b}_{B}>$. We concatenate the embeddings of \emph{timestamp, location, and behavior\_type} as the behavior's representation $\mathbf{e}_{b_i}=[\mathbf{e}_{b_i,itemid},\mathbf{e}_{b_i,timestamp},...,\mathbf{e}_{b_i,behavior\_type}]$. The intra-group behavior sequence can be expressed as $\mathbf{e}_b=[\mathbf{e}_{b_1}, \mathbf{e}_{b_2}, ..., \mathbf{e}_{b_B}]$.

Due to the ability to model behavior pairs from multiple perspectives, we apply Multi-Head Self Attention (MHSA) to capture the interest evolution. The MHSA can be expressed as follows:
\begin{equation}
    MHSA(\mathbf{e}_b)=concat(head_1,...,head_h)W^O,
\end{equation}
\begin{equation}
    head_i=Softmax(\frac{\mathbf{e}_bW^Q_i({\mathbf{e}_bW^K_i})^T}{\sqrt{d'}})\mathbf{e}_bW^V_i,
\end{equation}
where h is the number of heads, $W^Q_i, W^K_i, W^V_i \in R^{d \times d}$, $W^O \in R^{d \times d}$. $d$ is the dimension of the input. Then, mean pooling is taken to process the $MHSA(\mathbf{e_b})$ and acquire the aggregated attribute as Eq.~(\ref{eq:agg_attr}).
\begin{equation}
    \label{eq:agg_attr}
    \mathbf{dyna}_s=mean\_pool(MHSA(\mathbf{e}_b)).
\end{equation}
where $\mathbf{dyna_s}$ is the dynamic interest evolution of the $s$-th group.

\subsubsection{Top-k Index Search}
The maximum timestamp within a group is critical for characterizing group-level interests, as it indicates the recency of user engagement with the group—directly a key determinant of the group’s selection probability. Specifically, if a user has a recent interaction within a group, the likelihood of this group being prioritized increases. When groups are selected via top-k index search based on maximum timestamps, all behaviors within the selected groups are activated. These groups then undergo representation refinement and are fed into the Temporal Group Evolution Transformer (TGET). This process can be formally expressed as: Eq.~(\ref{eq:topk_search}).
\begin{equation}
    \label{eq:topk_search}
    \mathbf{G}= \sigma_{topk}([\mathbf{g}_{1},\mathbf{g}_{2},...,\mathbf{g}_{n}], \mathbf{max\_time}) 
\end{equation} 
\begin{equation}
        \mathbf{g}_s= concat[\mathbf{dyna}_s, \mathbf{stat}_s]
\end{equation}
where $\mathbf{g}_s$ denotes the representation of the $s$-th group, and $\mathbf{G}$ refers to the top-$k$ groups selected via a time-based search that prioritizes the maximum timestamp.

\subsection{Temporal Group Evolution Transformer Module}
\begin{figure}
    \centering    
    \includegraphics[width=1.0\linewidth]{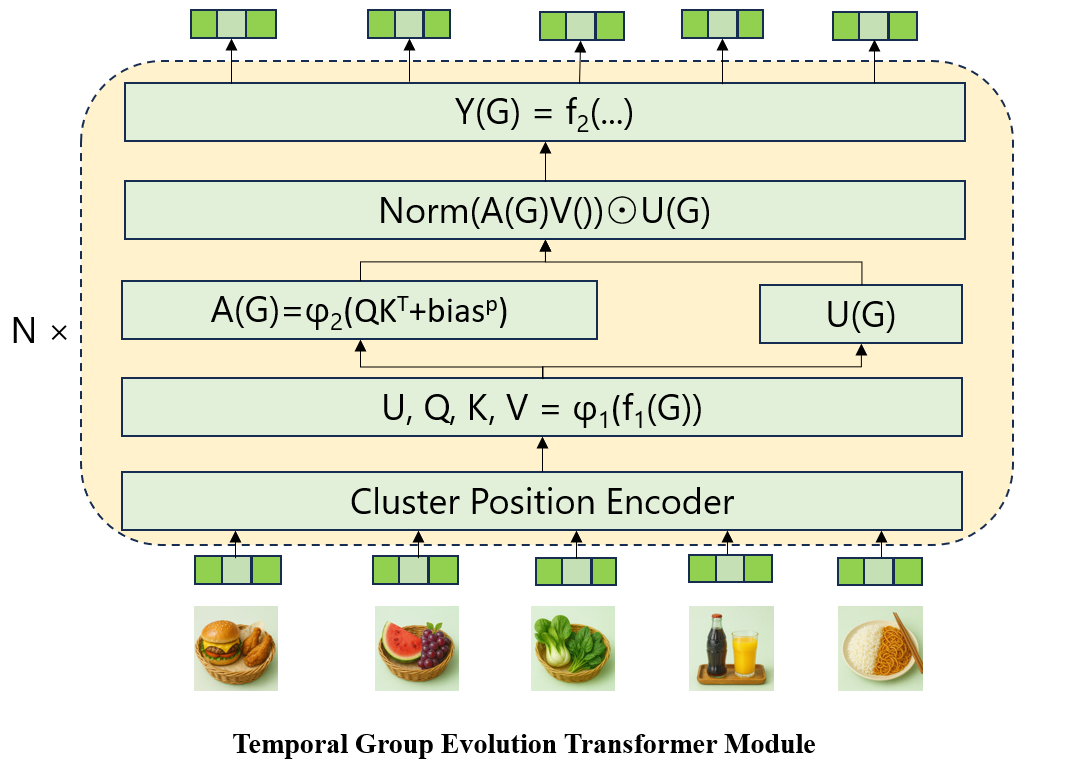}
    \caption{A stack of N Hierarchical Sequential Transduction Units (HSTU) is incorporated to model the temporal evolution of groups.}
    \label{fig:TGET}
\end{figure}
After extracting group representations, we employ multi-layer inter-group transformers on temporally ordered groups to capture the evolutionary patterns of users' group-based interests. Notably, we select Hierarchical Sequential Transduction Units (HSTU)\cite{zhai2024actions} as the transformer component block. As a core component of industrial long-range models (LRM), HSTU plays a critical role: it enables in-depth exploration of the temporal evolution and mutual influence of interest groups. This interaction mechanism enhances the model's comprehension of users' interest structures, facilitating more accurate representation of their preferences derived from historical behavior patterns. This process can be formally expressed as:
\begin{equation}
    \label{eq:qkvu}
    U(G), V (G), Q(G), K(G) = \text{Split}(\varphi_1(f_1(G)))
\end{equation}
\begin{equation}
    \label{eq:qk}
    A(G)V(G) = \varphi_2(Q(G)K(G) + bias^p)V(G) 
\end{equation}
\begin{equation}
    \label{eq:u}
    Y(G) = f_2 (\text{Norm}(A(G)V(G)) \mathbf{\odot} U(G)) 
\end{equation}
where $f_i(X)$ denotes an MLP; we use one linear layer,
$f_i(G) = W_i(G) + b_i$ for $f_1$ and $f_2$ to reduce compute
complexity and further batches computations for queries
Q(G), keys K(G), values V(G), and gating weights U(G)
with a fused kernel; $\varphi_1$ and $\varphi_2$ denote nonlinearity, for
both of which we use SiLU; Norm is layer norm; and $bias^p$ denotes relative cluster bias.
\subsection{Candidate-Aware Group Attention Module}
\begin{figure}
    \centering    
    \includegraphics[width=1.0\linewidth]{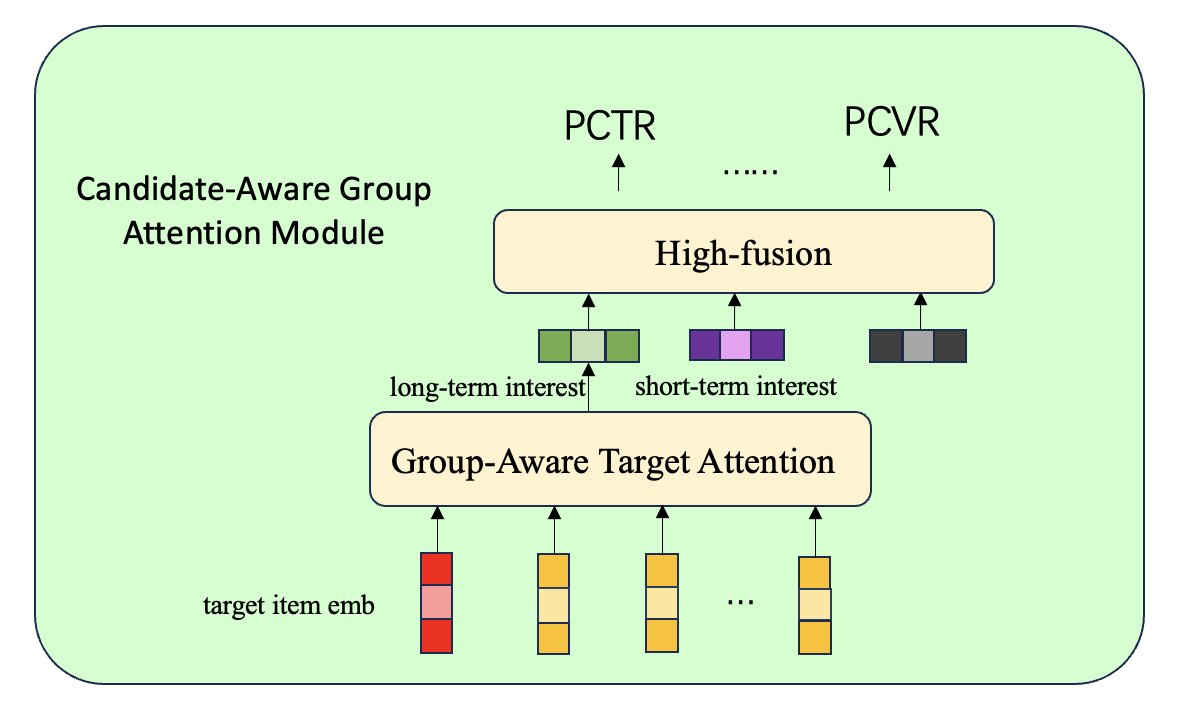}
    \caption{The high-level fusion input comprises long-term interest representations, short-term interest signals, and additional auxiliary features.}
    \label{fig:CAGAM}
\end{figure}
To derive candidate-aware representations aligned with long-term group interests, we refine the target attention mechanism by computing attention scores between target items and groups, thereby identifying candidate-specific interest signals. This process can be encoded as:
\begin{equation}
    \label{eq:s_i}
\mathbf{r}_{long} = \frac{\exp\left(\text{sim}(\mathbf{W}_t \mathbf{e}, \mathbf{W}_g \mathbf{g'}_s)\right)}{\sum_{s'=1}^{S} \exp\left(\text{sim}(\mathbf{W}_t \mathbf{e}, \mathbf{W}_g \mathbf{g'}_{s'})\right)}
\end{equation}
Where $\mathbf{e^t}$ is the embedding of target item. 
\(\mathbf{g'}_s \in \mathbf{R}^d\) represents the \(s\)-th group evoluted embedding;
\(\mathbf{W}_t, \mathbf{W}_g \in \mathbf{R}^{d \times d}
\) are learnable projection matrices;
$\mathbf{r}_{final}$ represents the user's long-term interest representation. 
\subsection{Caching The Intermediate Representation}
Since the output of TGETM is irrelevant to the candidate item, we can precompute the representation of each user's long behavior sequence and cache them in a key-value database such as Redis~\cite{carlson2013redis} after finishing training. When serving online, we can get the output from the key-value database based on user\_id rather than computing online which is computationally expensive. Thus, we can launch DMGIN successfully to meet the strict latency requirement. 


\section{Experiment Setup}
\subsection{Datasets}
Experiments are conducted on both industrial and public datasets. The statics values of each dataset are shown in Table~\ref{tab:data_stats}.

\textbf{Industry} is the CTR dataset collected from our online LBS platform. The last 30 days' logs are used for training and samples of the following day are for testing. To exploit abundant behavior information and multimodal content, we collect various historical behaviors of each user from the past 2 years. The maximum length of the full lifelong behavior sequence is 10000. There are 76 types of behaviors in our RS including click, add-to-cart, add-to-favorite, browse-dishes, view-comments, order and so on. 

\textbf{Amazon}~\cite{mcauley2015image} is a widely used public benchmark for sequential recommendation. We use the Grocery and Gourmet Food subset, which contains 287209 distinct products, each annotated with rich meta-features such as hierarchical category labels, brand, package size, price and nutrition attributes. The corpus comprises 5074160 time-stamped reviews, where every review is accompanied by a set of fine-grained post-click actions. For this dataset, the maximum historical interaction length per user is 100. We therefore partition each user’s history into a short and a long segment: the 10 most recent actions form the short sequence, while the latest 90 comprise the long sequence.
  
\begin{table}[h]
\caption{Statistics of datasets.}
\centering
\label{tab:data_stats}
\begin{tabular}{lc|c|c|c|c}
\toprule
\multicolumn{2}{c|}{Datasets}   & \multicolumn{1}{c}{\#Users} & \multicolumn{1}{c}{\#Items} & \multicolumn{1}{c}{\#Fields} & \multicolumn{1}{c}{\#Instances} \\
\midrule
\multicolumn{2}{c|}{Amazon}           & \multicolumn{1}{c}{1357K}             & \multicolumn{1}{c}{474K}            & \multicolumn{1}{c}{19}               & \multicolumn{1}{c}{72.17M}  \\
\multicolumn{2}{c|}{Industry}           & \multicolumn{1}{c}{400M}             & \multicolumn{1}{c}{5M}            & \multicolumn{1}{c}{317}               & \multicolumn{1}{c}{6.6B}  \\
\bottomrule 
\end{tabular}
\end{table}

\subsection{Evaluation Metric}
Two widely used metrics, AUC~\cite{bishop2006pattern} and Group AUC (GAUC), are employed. AUC (Area Under the ROC Curve) measures ranking accuracy, with a higher value indicating better performance; GAUC extends this by accounting for group-level biases, providing a more robust evaluation across user groups.

 \subsection{Baselines}
We select baseline models for comparison from three perspectives. First, we include methods focusing on short behavior sequence modeling, such as \textbf{DIN}~\cite{zhou2018deep}, \textbf{DIEN}~\cite{zhou2019deep}, and \textbf{DSIN}~\cite{feng2019deep}, which extract user interests from short click behavior sequences. Second, we incorporate recent two-stage retrieval-refinement models: \textbf{SIM}~\cite{pi2020search}, \textbf{ETA}~\cite{chen2022efficient}, \textbf{SDIM}~\cite{cao2022sampling}, and \textbf{TWIN}~\cite{chang2023twin}, which enable efficient modeling of long sequences through a search paradigm. Third, \textbf{TWIN V2}~\cite{si2024twin} focuses on full lifelong behavior sequence modeling. Additionally, \textbf{DSIN Full} and \textbf{DIN Full}, with their hierarchical attention structures, are also included as baselines for direct modeling of full lifelong behavior sequences.

\subsection{Implementation Details}
For SIM, ETA, SDIM, and TWIN,  we retrieve the top 50 most candidate-relevant behaviors into the second interest extraction stage. We implement DMGIN with Tensorflow. For the industry dataset, the embedding size is $16$, clusters of k-kmeans is $10000$ and the learning rate is $5e-4$. We train the model using eight $80G$ $A100$ GPUs with the batch size 1024 of a single card. For the Taobao dataset, we set the embedding size to be $18$, the learning rate to be $1e-3$, and use a single $80$ $A100$ for training with batch size $1024$. We use Adam~\cite{kingma2014adam} as the optimizer for both datasets. We run all experiments five times and report the average result. 

\section{Experiment Results}
\subsection{Overall Performance}

\begin{table}[tb!]
\centering
\caption{Performance of all methods on both datasets. The best result is in boldface and the second best is underlined. * indicates that the superiority to the best baseline is statistically significant at 0.01 level.}
\label{tab:main_result}
\begin{tabular}{cl|c|c|c|c}
\toprule
& & \multicolumn{2}{c}{Industry} &\multicolumn{2}{c}{Amazon} \\
& & \multicolumn{1}{c}{AUC} &\multicolumn{1}{c}{GAUC} &\multicolumn{1}{c}{AUC} &\multicolumn{1}{c}{GAUC} \\
\midrule
\multicolumn{2}{c|}{DIN}        & \multicolumn{1}{c}{0.7011}                & \multicolumn{1}{c|}{0.6391}                & \multicolumn{1}{c}{0.7245}                & \multicolumn{1}{c}{0.6753} \\

\multicolumn{2}{c|}{DSIN}        & \multicolumn{1}{c}{0.7029}                & \multicolumn{1}{c|}{0.6391}                & \multicolumn{1}{c}{0.7284}                & \multicolumn{1}{c}{0.6781} \\

\multicolumn{2}{c|}{DIEN}       & \multicolumn{1}{c}{0.7035}                & \multicolumn{1}{c|}{0.6406}                & \multicolumn{1}{c}{0.7303}                & \multicolumn{1}{c}{0.6796} \\

\multicolumn{2}{c|}{SIM}    & \multicolumn{1}{c}{0.7043}                & \multicolumn{1}{c|}{0.6412}                & \multicolumn{1}{c}{0.7468}                & \multicolumn{1}{c}{0.6832} \\

\multicolumn{2}{c|}{ETA}      & \multicolumn{1}{c}{0.7017}                & \multicolumn{1}{c|}{0.6381}                & \multicolumn{1}{c}{0.7452}                & \multicolumn{1}{c}{0.6826} \\

\multicolumn{2}{c|}{TWIN}       & \multicolumn{1}{c}{{0.7051}}    & \multicolumn{1}{c|}{{0.6396}}    & \multicolumn{1}{c}{0.7480}    & \multicolumn{1}{c}{0.6851}  \\ 

\multicolumn{2}{c|}{SDIM}       & \multicolumn{1}{c}{0.7073}                & \multicolumn{1}{c|}{0.6409}                & \multicolumn{1}{c}{0.7497}                & \multicolumn{1}{c}{0.6844} \\ 

\multicolumn{2}{c|}{DIN Middle}       & \multicolumn{1}{c}{{0.7077}}   & \multicolumn{1}{c|}{{0.6420}}   & \multicolumn{1}{c}{0.7521}   & \multicolumn{1}{c}{0.6867} \\ 

\multicolumn{2}{c|}{TWIN V2}       & \multicolumn{1}{c}{{0.7078}}    & \multicolumn{1}{c|}{\underline{0.6421}}    & \multicolumn{1}{c}{0.7526}    & \multicolumn{1}{c}{0.6878}  \\ 

\multicolumn{2}{c|}{DIN Full}       & \multicolumn{1}{c}{{0.7087}}   & \multicolumn{1}{c|}{\underline{0.6421}}   & \multicolumn{1}{c}{0.7634}   & \multicolumn{1}{c}{0.6934} \\ 

\multicolumn{2}{c|}{DSIN Full}       & \multicolumn{1}{c}{\underline{0.7121}}   & \multicolumn{1}{c|}{\underline{0.6421}}   & \multicolumn{1}{c}{\underline{0.7663}}   & \multicolumn{1}{c}{\underline{0.6955}} \\ 
\midrule

\multicolumn{2}{c|}{DMGIN}       & \multicolumn{1}{c}{\textbf{0.7161}}   & \multicolumn{1}{c|}{\textbf{0.6458}}   & \multicolumn{1}{c}{\textbf{0.7747}}   & \multicolumn{1}{c}{\textbf{0.7022}} \\ 

\bottomrule 
\end{tabular}
\end{table}
Table ~\ref{tab:main_result} summarizes the performance of every model. DMGIN obtains the best performance in both the Industry and Amazon datasets, which shows the effectiveness of DMGIN. There are some insightful findings from
the results:

The proposed DMGIN reaches the best performanceon both datasets, DMGIN in terms of sequence grouping, DMGIN leverages Cross-Modal Representation Learning Module (CMRLM)and Interest- DrivenEntity Clustering Module (IDECM) to reorganize users' complete lifelong post-click behavior sequences. This approach not only achieves more effective grouping but also avoids the computational burden of directly introducing multimodal embeddings, resulting in negligible additional overhead—a critical advantage for industrial-scale deployment.
To address potential information loss caused by grouping, DMGIN introduce Intra-Group Enhancement Module (IGIEM): Within each group, it combines interest statistics with intra-group transformers, enabling fine-grained capture of group-specific traits and mitigating the loss of local behavioral patterns.
Across groups, it applies Temporal Group Group Evolution Transformer Module (TGETM) to temporally ordered groups, effectively modeling the evolutionary dynamics of user interests at the group level, thus preserving the global sequence structure.

DSIN's better performance than DIN vividly illustrates the crucial role of interaction in extracting users' session interests. As data complexity rises, DIN has limits, while DSIN's enhanced interaction dissects session interests more precisely. This underlines the need for stronger interaction. The Intra-Group Behavior Evolution is thus introduced. It aims to supercharge interaction via  self-attention architectures, unearthing deeper insights and more accurate interest representations, enhancing group interest representation.

The performance boost from the two-step long behavior sequence modeling process highlights its significance. SIM, SDIM, TWIN, and TWIN V2 outperform DIN, DIEN, and DSIN. Short sequences offer a limited view, while long ones comprehensively mirror users' complex, evolving interests and behaviors. Leveraging the full long sequence, as we plan, is crucial for deeper user understanding and application optimization.

\subsection{Ablation Study}

\subsubsection{Intra-Group Interest Enhancement Module}
\begin{table}[tb!]
\centering
\caption{Results of integrating Intra-Group Interest Enhancement Module successively.}
\vspace{-0.4cm}
\label{tab:ablation_component}
\begin{tabular}{cl|c|c|c|c}
\toprule
& & \multicolumn{2}{c}{Industry} &\multicolumn{2}{c}{Amazon} \\
& & \multicolumn{1}{c}{AUC} &\multicolumn{1}{c}{GAUC} &\multicolumn{1}{c}{AUC} &\multicolumn{1}{c}{GAUC} \\
\midrule
\multicolumn{2}{c|}{TWIN V2}                       & \multicolumn{1}{c}{0.7078}            & \multicolumn{1}{c|}{0.6421}               & \multicolumn{1}{c}{0.7526}            & \multicolumn{1}{c}{0.6878} \\ 
\multicolumn{2}{c|}{DSIN Full}                       & \multicolumn{1}{c}{0.7121}            & \multicolumn{1}{c|}{0.6421}               & \multicolumn{1}{c}{0.7663}            & \multicolumn{1}{c}{0.6955} \\ 

\midrule
\multicolumn{2}{c|}{DMGIN-simple}                & \multicolumn{1}{c}{\textbf{0.7134}}           & \multicolumn{1}{c|}{\textbf{0.6440}}            & \multicolumn{1}{c}{\textbf{0.7701}}            & \multicolumn{1}{c}{\textbf{0.6986}} \\ 

\midrule
\multicolumn{2}{c|}{-Interest Statistical}    & \multicolumn{1}{c}{0.7129}            & \multicolumn{1}{c|}{0.6435}               & \multicolumn{1}{c}{0.7679}            & \multicolumn{1}{c}{0.6977} \\
\multicolumn{2}{c|}{-Behavior Evoltion}     & \multicolumn{1}{c}{0.7110}            & \multicolumn{1}{c|}{0.6435}               & \multicolumn{1}{c}{0.7685}            & \multicolumn{1}{c}{0.6984} \\

\bottomrule 
\end{tabular}
\end{table}
Prior to introducing the Temporal Group Evolution Transformer Module, we performed an ablation study on the Intra-Group Interest Enhancement Module to quantify the contribution of its two principal subcomponents, namely interest-statistics features and intra-group behavior-evolution modeling, to overall performance. The complete IGIEM already allows DMGIN-simple to outperform the strong baselines TWIN V2 and DSIN Full on both the Industry and Amazon datasets. Excluding the interest-statistics branch leads to only a minor decline across all metrics, whereas removing the behavior-evolution branch produces a markedly larger drop; on the Industry dataset, for example, the reduction in AUC is nearly five times greater than that observed when the statistics branch is removed. These findings indicate that static statistical cues are insufficient to capture group interests comprehensively, and that modeling the dynamic evolution of behavior is the main driver of the observed performance gains.

\subsubsection{Temporal Group Evolution Transformer Module}
\begin{figure}
    \centering    
    \includegraphics[width=1.0\linewidth]{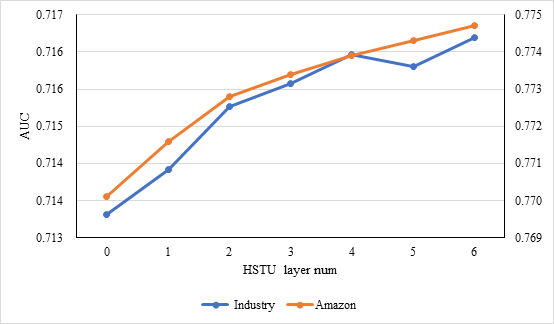}
    \caption{The Scaling law of Temporal Group Evolution Transformer Module}
    \label{fig:CAGA}
\end{figure}
We conduct ablation experiments to investigate how model depth and embedding dimensionality affect DMGIN’s performance. This analysis focuses on the Temporal Group Evolution Transformer Module(TGETM), an important DMGIN component composed of stacked Hierarchical Self-attention Transduction Units (HSTUs) that capture complex temporal-group interactions. By growing the TGETM depth from 1 to 6 layers, we found a strictly monotonic accuracy rise: each added block deepens temporal-group reasoning and orchestrates more intricate cross-interest interactions, lifting AUC/GAUC on both Amazon and Industrial benchmarks without any hint of saturation. On Amazon, for example, six stacked HSTUs increase AUC by about 0.3\%–0.5\% over a single layer, with a comparable gain on the Industrial set. These results show that richer interaction pathways inside TGETM are still being tapped as we stack more HSTUs, implying the model’s capacity limit has not yet been reached and that even deeper configurations could push performance further if computation allows.

\subsection{A/B Test on Performance and Cost}
Finally, we evaluate DMGIN in a live production environment to verify that its offline improvements translate to real-world gains and to assess its system overhead. We deployed DMGIN in industrail’s large-scale advertising system and conducted an online A/B test against the incumbent production model (a two-stage long-sequence CTR model) over a period of several weeks. In this test, a small portion of user traffic was served by DMGIN while the majority continued to use the baseline model for control. The results demonstrate substantial online benefits: DMGIN improved the click-through rate (CTR) by 4.7\% and the revenue-per-mille (RPM) by 2.3\% relative to the baseline. These increases in key business metrics validate the practical value of our approach, as even single-percentage-point CTR lifts are considered highly significant in industrial recommendation systems.

In terms of system efficiency, deploying DMGIN incurred only a modest increase in resource usage, which is well justified by its gains. The model’s parameter storage footprint grew from roughly 2.8 GB (for the previous baseline) to about 4.0 GB for DMGIN, owing to the additional cluster embeddings and HSTU layers. The average inference latency also rose from about 4–5 ms to 7–8 ms per request due to the deeper architecture. This ~2–3 millisecond latency overhead is relatively small and has minimal impact on user experience, especially considering that DMGIN processes the full behavior sequence rather than a truncated subset. Overall, the cost-performance trade-off of DMGIN is favorable: the slight increase in memory and computation is negligible in the context of a large-scale system, whereas the online CTR and RPM gains are substantial. These outcomes highlight that DMGIN is not only effective but also deployable in an industrial setting. By preserving long-term user history in a compressed form and leveraging multimodal knowledge with manageable overhead, DMGIN provides a practical, scalable solution for modern recommendation services.

\section{Conclusion}

In this paper, we propose the DMGIN for full long user behavior sequence modeling in the CTR prediction task. Aims at extracting fine-grained comprehensive unbiased interest and psychological decision interest to achieve a deep understanding of the user's preference. To the best of our knowledge, DMGIN is the first to achieve efficient end-to-end full long user behavior sequence modeling.


\appendix

\bibliography{aaai2026}


\end{document}